\documentclass[twoside,fleqn]{article}
\usepackage{epsfig}
\usepackage{espcrc2}

\def\vev#1{\langle #1 \rangle}

\title{Super-heavy Quarkonia as a probe of the QCD Vacuum}

\author{J.~Fingberg, University of Wuppertal, 42097 Wuppertal, Germany}

\begin{document}
\begin{abstract}
We investigate the effect of a non-perturbative gluon
condensate on the spectrum of bound states of extremely
heavy quarks using Non-Relativistic-QCD (NRQCD).
Simulations are done with an improved gluon action on
small lattices. We compare our numerical results with
perturbative calculations to test the consistency.
\end{abstract}
\maketitle
The phenomenon of gluon condensation is under vital discussion since more
than fifteen years now \cite{Gromes91}.
The gluon condensate is an important quantity not only in the sum rule approach
but also in high-energy hadron scattering and phenomenological models of
confinement.
Despite many indications of a non-trivial vacuum the quantitative picture is
still vague \cite{cond}.
Quarkonia are the simplest hadronic systems that allow to test the
vacuum structure of QCD. Heavy quarks can be treated as non-relativistic
objects in a gluonic vacuum making them tangible to numerical simulations
and perturbative calculations.
\setlength{\unitlength}{0.30mm}
\begin{center}
\begin{picture}(100, 60)
\linethickness{0.72pt}
\put(  0,  0){\line(0,1){ 60}}
\put(  0,  0){\line(1,0){100}}
\put(100, 60){\line(0,-1){ 60}}
\put(100, 60){\line(-1,0){100}}
\put( 10,  0){\line(0,1){ 60}}
\put( 20,  0){\line(0,1){ 60}}
\put( 30,  0){\line(0,1){ 60}}
\put( 40,  0){\line(0,1){ 60}}
\put( 50,  0){\line(0,1){ 60}}
\put( 60,  0){\line(0,1){ 60}}
\put( 70,  0){\line(0,1){ 60}}
\put( 80,  0){\line(0,1){ 60}}
\put( 90,  0){\line(0,1){ 60}}
\put(  0, 10){\line(1,0){100}}
\put(  0, 20){\line(1,0){100}}
\put(  0, 30){\line(1,0){100}}
\put(  0, 40){\line(1,0){100}}
\put(  0, 50){\line(1,0){100}}
\put(  0, 60){\line(1,0){100}}
\put( 40, 30){\circle*{10}}
\put( 60, 30){\circle*{10}}
\linethickness{2.0pt}
\put( 40, 30){\line(1,0){20}}
\linethickness{1.00pt}
\put( 50, 30){\circle {36}}
\linethickness{0.72pt}
\put( 21, 36){\makebox(0,0)[l]{$Q$}}
\put( 71, 37){\makebox(0,0)[l]{$\bar Q$}}
\end{picture}
\end{center}
In this study we compare the quark mass dependence of quarkonium 
level splittings
from simulations using NRQCD \cite{Lepage} to analytic results.
The effect of the gluon condensate can be calculated perturbatively
assuming that super-heavy quarkonia are almost Coulombic:
$E_{nl} = 2 m_Q + E_{nl}^{Balmer} + \delta E_{nl}$.
A non-vanishing gluon condensate \cite{Zakharov79}
$\vev{E^2}=-1/4\vev{0 | G_{\mu\nu}^a G_{\mu\nu}^a | 0}\neq 0$
gives rise to a level shift $\delta E_{nl}$ by quadratic Stark splitting.
The simplest ansatz is a constant chromo-electric
field \cite{Voloshin79,Leutwyler81}. In this case the shift
depends on the principal quantum number and
the quark mass in the form $\delta E_{nl}\propto n^6/m_Q^3$.
A finite gluon correlation length \cite{Balitzky85,Gromes82}
($\lambda_G \simeq 0.22\mbox{~fm}$ \cite{DiGiacomo92})
weakens this dependence: $\delta E_{nl}\propto n^4/m_Q^2$.
There are two length scales in the system,
$\lambda_G$, and the quark correlation length, $\lambda_Q^{nl}$, thus
two limiting cases can be considered \cite{MD87}:
the non-potential case $\lambda_G>>\lambda_Q^{nl}$ and
the potential case $\lambda_G<<\lambda_Q^{nl}$.
The first case reduces to Leutwyler's result,
$\Delta E_{21}=E_{20}-E_{10}=c_1 m_Q/n^2+c_2 n^6/m_Q^3$ as $m_Q\to\infty$.
For the second case one finds $\Delta E_{21}=c_3 m_Q/n^2+c_4 n^4/m_Q^2$.
The hyperfine splitting of the ground state can be parametrized
in the same form.

Our simulations are done with a Symanzik and tadpole improved gauge action
with a plaquette and a rectangle term
\[
S[U]=\beta\left[\sum_{pl}(1-U_{pl})-
     \frac{1}{20u_0^2}\sum_{rt}(1-U_{rt})\right] ~~,
\]
enabling us to use a small lattice of size $16^4$.
The quark propagators are computed with the NRQCD evolution equation:
\[
G_{t+1}=(1-\frac{aH_0}{2n})^n~U^\dagger_4
        (1-\frac{aH_0}{2n})^n(1-a\delta H)~G_t.
\]
The Hamiltonian includes all
spin-independent interactions to ${\cal O}(v^2)$ and
spin-dependent interactions to ${\cal O}(v^4)$.
Difference operators are improved to avoid large discretization errors.
We generate $\approx300$ decorrelated configurations
at $\beta=9.17$ and determine the meson spectrum for S and P states.
To set the scale we measure the heavy quark potential
for on axis and $\sqrt{2}$ separations from Wilson loops
and determine the lattice spacing from the force \cite{Sommer94},
$F(R)={\partial V(R)}/{\partial R}$.
A comparison of dimensionless quantities $r_0^2~F(r_0)$
on the lattice and
using both Eichten's phenomenological potential,
\[
 V(r)=-{\kappa}/{r}+\sigma r~,~~~
 \sqrt{\sigma}=427\mbox{~MeV,~~}\kappa=0.52
\]
and L\"uscher's bosonic string model, $\kappa_L=\pi/12$,
shows good agreement of the sting model parameterization
with the lattice data. A fit shown in fig.1 gives the results
$r_0=0.34\mbox{~fm}$ and $a^{-1}=4.6(1)\mbox{~GeV}$.
\begin{figure}[htb]
          \epsfig{file=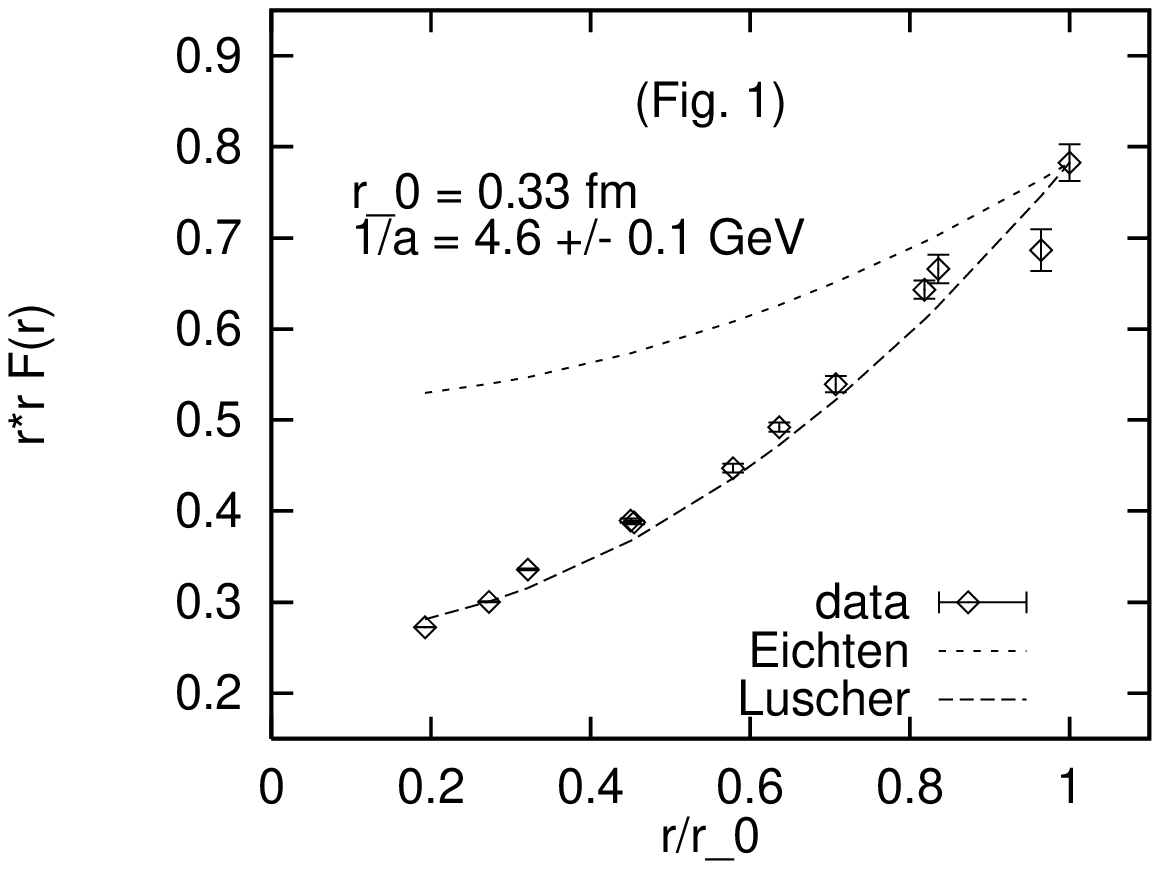,%
                  width=0.72\linewidth,%
                  angle=270}
\end{figure}
Meson propagators and effective masses are measured
for 4 values of the bare quark mass, $m_Qa=1.5, 2, 3, 4$,
using Coulomb wave functions.
The kinetic mass is extracted from the dispersion relation,
$m(\vec p)=m_{sim}+{|\vec p~|^2}/{2m_{kin}}-{|\vec p~|^4}/{8m_2^3}$,
by a fit to ${}^1S_0$ finite momentum meson masses.
\begin{figure}[htb]
          \epsfig{file=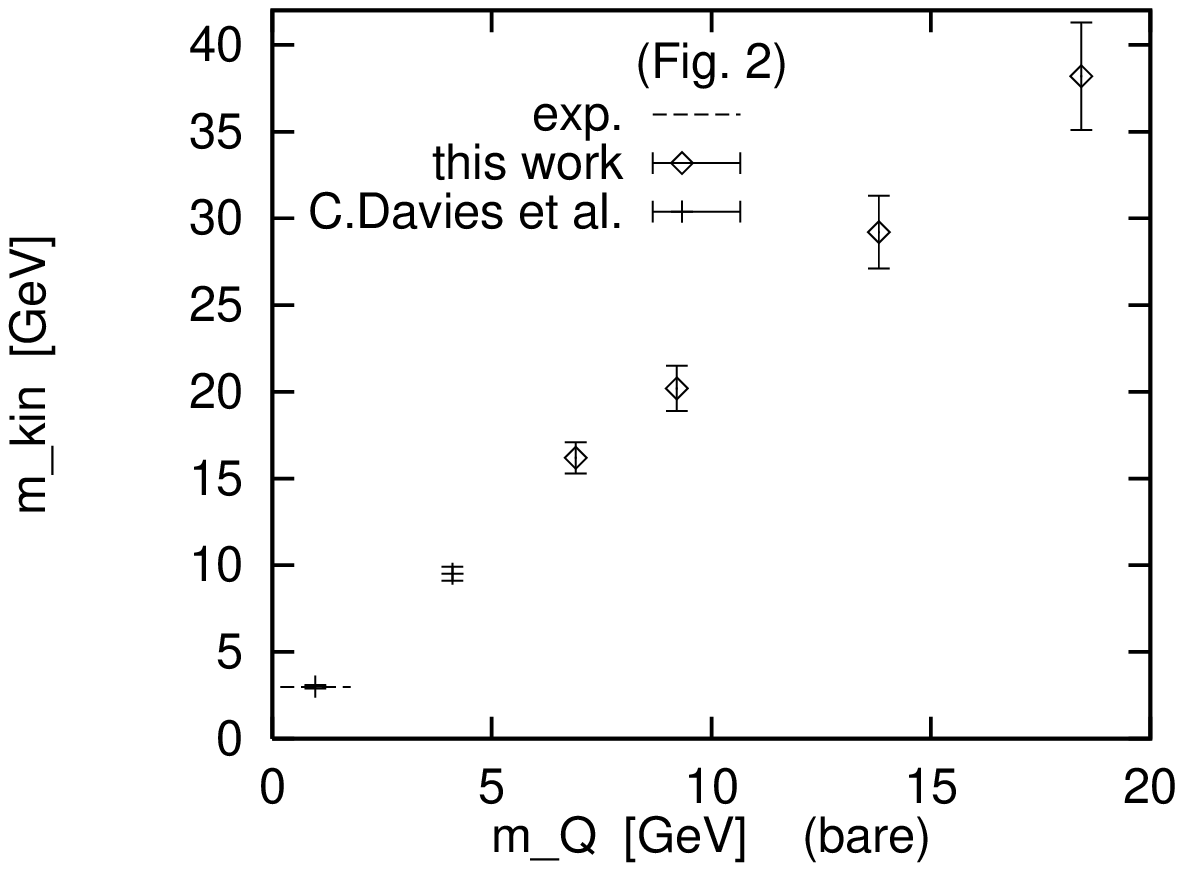,%
                  width=0.72\linewidth,%
                  angle=270}
\end{figure}
In fig.2 we see that for low $m_Q$ the results nicely extrapolate
to the $c\bar c$ and $b\bar b$ points of \cite{CTHD}.
\begin{figure}[htb]
          \epsfig{file=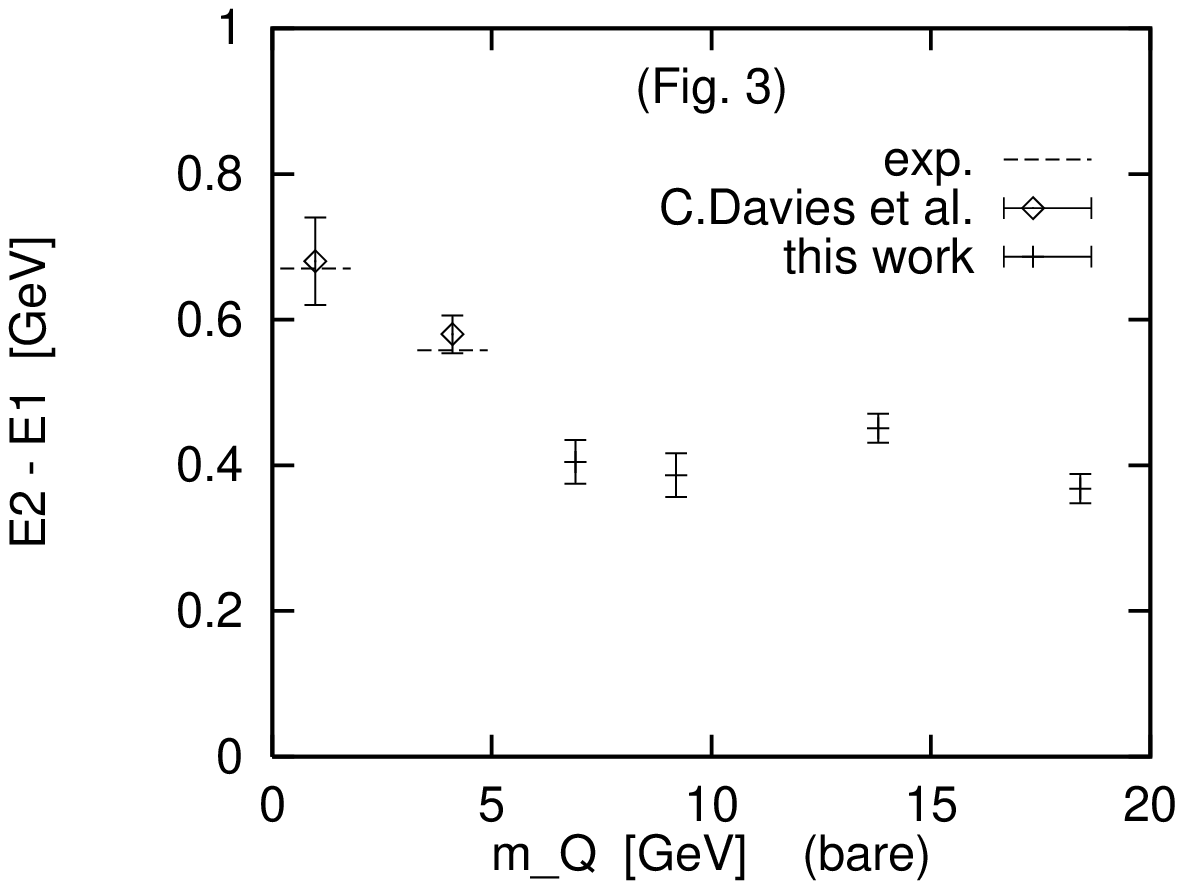,%
                  width=0.72\linewidth,%
                  angle=270}
          \epsfig{file=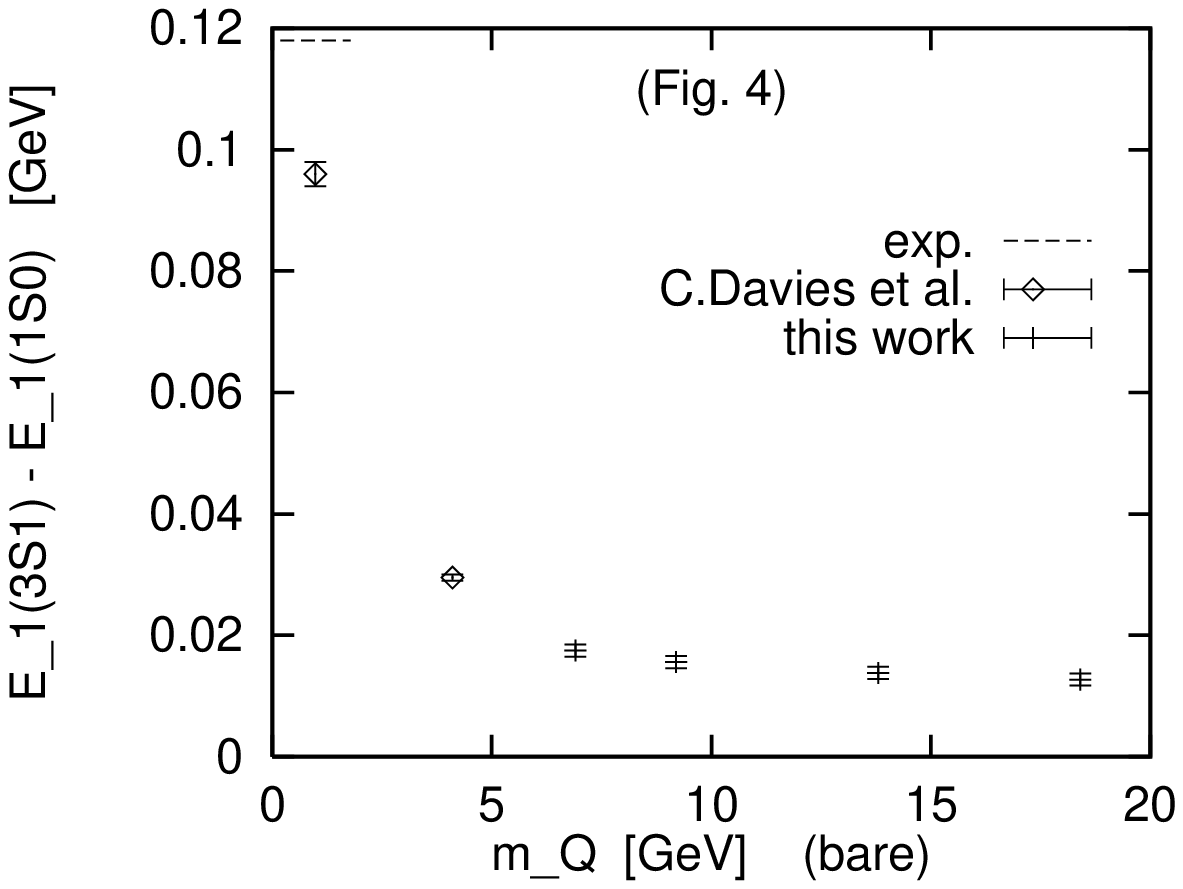,%
                  width=0.72\linewidth,%
                  angle=270}
\end{figure}
The splitting $\Delta E_{21}=E_2({}^1S_0$)-$E_1({}^1S_0)$ plotted in fig.3
is determined from correlated multi-exponential fits.
The experimental values also indicated in fig.3
are 670 MeV for $c\bar c$ and 558 Mev for $b\bar b$.
The important observation is that the new data points continue
to decrease with increasing quark mass. The effect becomes
smaller with increasing quark mass and vanishes in the noise
for the two largest masses.
The hyperfine splitting can be determined very accurately
from a fit to the ratio of ${}^3S_1$
and ${}^1S_0$ meson propagators with a single exponential.
The experimental value for $\Delta E_{hfs}^{c\bar c}$ is 118 MeV.
We observe a steep decrease with quark mass which levels off at higher
values of $m_Q$.
From analytic calculations we expect a behaviour of
both splittings, $\Delta E_{21}$ and $\Delta E_{hfs}$,
proportional to the quark mass for a pure Coulomb potential.
A gluon condensate would modify this result to
$\Delta E=c_1 m_Q+c_2\vev{\alpha_sG^2}/m_Q^{\alpha}$, with an
exponent $\alpha=2~...~3$
depending on the relative magnitude of $\lambda_G$ and $\lambda_Q^{nl}$.
In fig.5 we see qualitative agreement between our data and analytic
predictions if the value of the gluon condensate is large enough.
The observed decrease of the splittings does not agree with
the expectation from a pure Coulomb potential.
The apparent contradiction is likely to prevail for any reasonable potential
model calculation \cite{pot}.
\begin{figure}[htb]
          \epsfig{file=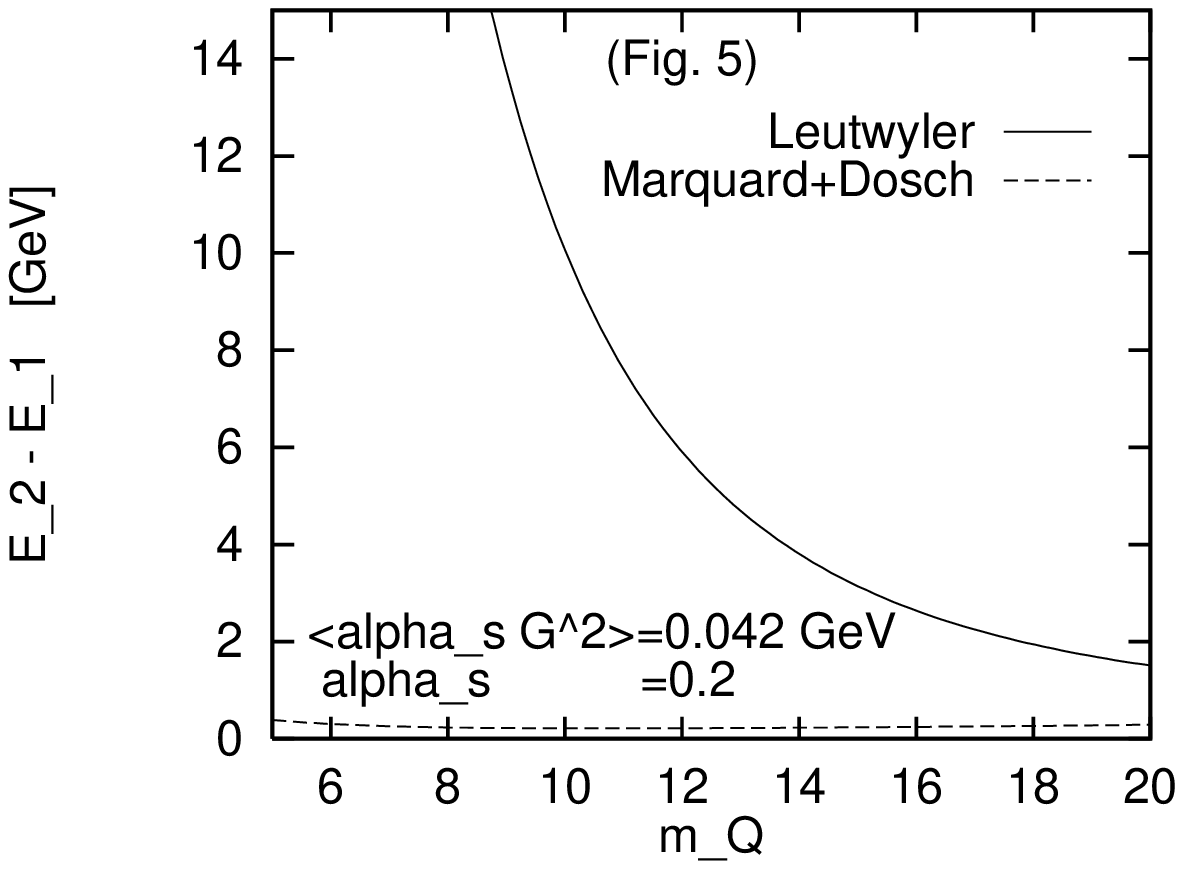,%
                  width=0.72\linewidth,%
                  angle=270}
          \epsfig{file=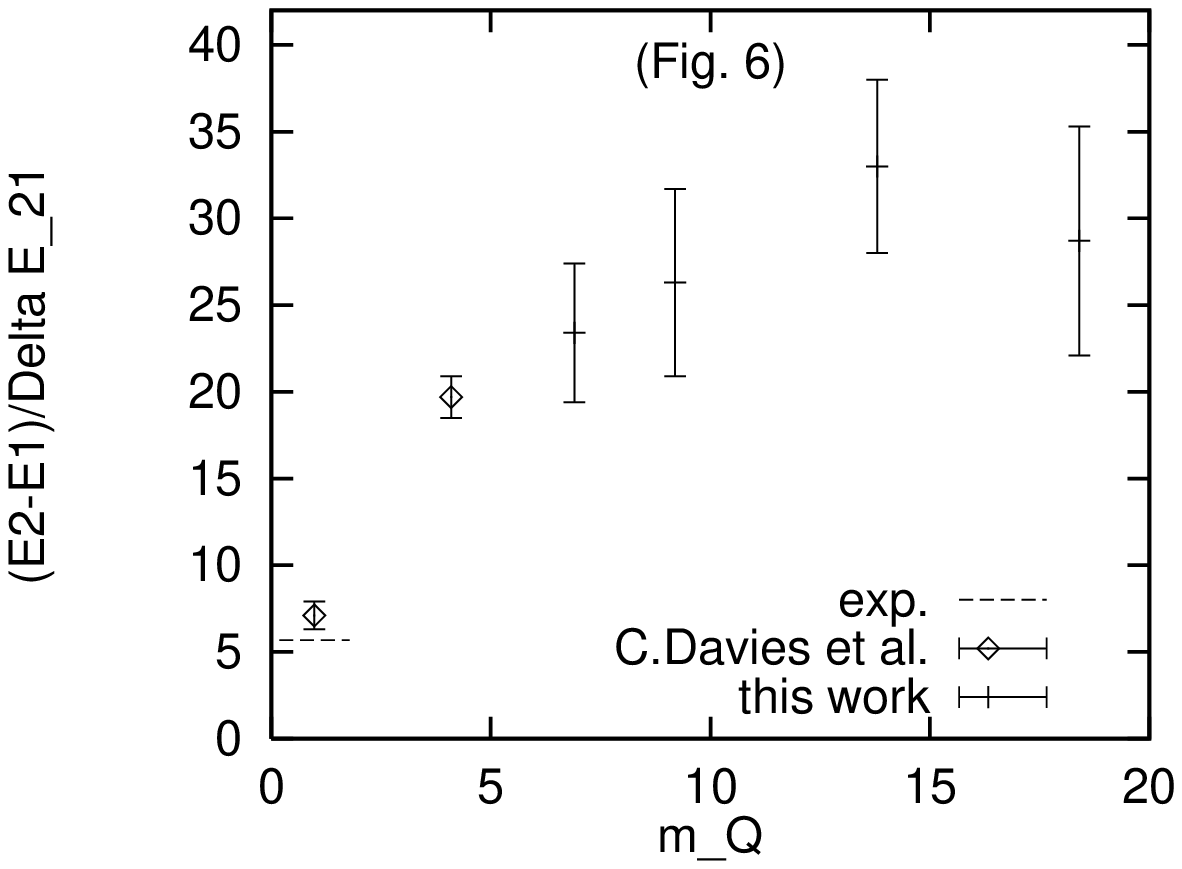,%
                  width=0.72\linewidth,%
                  angle=270}
\end{figure}
Finally the ratio of the ${}^1S_0$ and the
hyperfine splitting shows an increase at least up to $m_Q=15$~GeV while a
pure Coulomb potential would give a constant.
The experimental value is $r_{c\bar c}=\Delta E_{21}/\Delta E_{hfs}=5.7$.
\section*{SUMMARY}
We present results of a NRQCD analysis based on simulations with a
Symanzik and tadpole improved gauge action on a small lattice.
We find that $E_2({}^1S_0)-E_1({}^1S_0)$ and the hyperfine splitting
of the ground state decrease with increasing $m_Q$ in accord to
existing data for $c\bar c$ and $b\bar b$ systems at small $m_Q$.
The decrease is in qualitative agreement with model calculations
of gluon condensation if the gluon condensate is large enough.
However, present systems are still far from Coulombic.
A direct determination of $\vev{\alpha_s~G^2}$ is difficult and
we need even higher quark masses $m_Q>20~\mbox{GeV}$
and also higher excitations $n=3,4$ (in preparation).
A promising signal which requires extreme precision
is the S-P splitting and P-state hyperfine splitting
as noted in \cite{TY95}.
On the analytical side an improved perturbative calculation of the
corresponding matrix elements is desirable.
\section*{ACKNOWLEDGMENTS}
We thank Andrei Leonidov for inspiring this study,
Edwin Laermann for useful conversations and Burkhard Beinlich
for the initial version of the Quadrics program.
The computations were performed on the Quadrics Q4 and the
Connection Machine CM2 in Wuppertal.
J.F. is supported by a DFG fellowship.

\end{document}